\documentclass[aps,twocolumn,showpacs,prb]{revtex4}
\usepackage[dvips]{graphicx}

\begin{document}

\title{Distribution of localized states from fine
analysis of electron spin resonance spectra
of organic semiconductors: Physical meaning and methodology}
\author{Andrey S. Mishchenko$^{1,2}$, Hiroyuki Matsui$^{3}$, and 
Tatsuo Hasegawa$^3$}
\affiliation{
\mbox{$^1$Cross-Correlated Materials Research Group (CMRG), 
RIKEN Advanced Science Institute (ASI), Wako 351-0198, Japan} \\
\mbox{$^2$RRC ``Kurchatov Institute'' - 123182 - Moscow - Russia}\\
\mbox{$^3$National Institute of Advanced Industrial Science and Technology 
(AIST), AIST Tsukuba Central 4, Tsukuba 305-8562, Japan} \\
}
\date{\today}

\begin{abstract}
We develop an analytical method for the processing of electron 
spin resonance (ESR) spectra. The goal is to obtain the distributions of 
trapped carriers over both their degree of localization and their binding energy 
in semiconductor crystals or films composed of regularly
aligned organic molecules  [Phys. Rev. Lett. {\bf 104}, 056602 (2010)].      
Our method has two steps. 
We first carry out a fine analysis of the shape of the ESR spectra 
due to the trapped carriers; this reveals the distribution 
of the trap density of the states over the degree of localization.
This analysis is based on the reasonable assumption that the 
linewidth of the trapped carriers is predetermined by their degree 
of localization because of the hyperfine mechanism.  
We then transform the distribution over the 
degree of localization into a distribution over the binding energies.
The transformation uses the relationships between
the binding energies and the localization parameters of the trapped carriers.  
The particular relation for the system under study is obtained 
by the Holstein model for trapped polarons using a diagrammatic Monte
Carlo analysis.  
We illustrate the application of the method to 
pentacene organic thin-film transistors.     
\end{abstract}

\maketitle

\section{Introduction}

The electron spin resonance (ESR) technique offers a unique 
microscopic probe of carriers in semiconductors with unpaired spins.
It measures transitions between the quantum levels
$m_s = \pm 1/2$ in the presence of a magnetic field
\cite{Anderson,Feher,Altshuler,Sli10}.
The spectrum of the transition between two quantum levels
constitutes a $\delta$-function provided there is 
no external interference with the quantum levels 
of the spin system.   
However, the quantum states of carrier spins undergo 
a variety of interactions with the environment.
This interaction destroys the $\delta$-functional 
form of the spectrum and broadens it.   
The broadening of the ESR signal is a result of  
two fundamentally different contributions from
the medium.
The first is a decay of the quantum levels caused by 
interaction with the excitations of the environment. 
This mechanism leads to the Lorentzian shape 
of the spectral line.
The second contribution is a result of the  
inhomogeneities of the medium.
One example is the interaction with nuclear spins that is 
known as hyperfine interaction; here the inhomogeneities
are caused by the probability distribution of the nuclear spin 
moments.
In this case the spectroscopic signal is the sum 
of the contributions from spin systems located in
different surroundings.
The energy of signal in every surrounding is shifted by the local
magnetic field  depending on the environment so that 
the summed signal has an inhomogeneous shape.

As an example, electronic spin of cationic pentacene
molecule isolated in a solution exhibits 
inhomogeneous broadening of the ESR signal that arises because of 
hyperfine coupling with 14 proton nuclear spins.  
The signal is constituted of a series of individual lines
due to the hyperfine splitting.  
The envelope function of the signal is roughly reproduced 
by a Gaussian \cite{Bolton,BookESR,Lewis65}. 
This feature is consequence of  the central 
limit theorem (CLT).
The local magnetic field for each electronic spin is caused 
by interaction with 14 proton nuclear spins and the 
random energy shifts of respective ESR signals
are inevitably spread in accordance with Gaussian distributions, 
as a result of the independent nature of respective 
nuclear spin orientations.

We note that the anisotropic values of $g$-factors 
are averaged out 
(or narrowed motionally) by the rotation motion of the molecules
in solution. 
In contrast, solid-state organic molecular crystals exhibit 
ESR spectra composed of individual lines that are broadened 
due to the faster decay rate. Then, there might be the case 
that the individual lines have to become unresolved and the 
resulting ESR spectrum has to be very close to Gaussian 
envelope.

Recent major developments in the field of organic thin-film 
transistors (TFTs) allow high-precision field-induced ESR 
measurement (which is referred in the following simply as 
ESR) for the carriers in semiconductor crystals or films 
composed of regularly aligned organic molecules. 
In the measurements, carries are doped without introducing 
any randomness by using the field-induced technique. 
The ESR signal in organic TFTs was first measured and analyzed 
in the groundbreaking study by Marumoto and his coauthors
\cite{Maru06,Maru05}. 
The ESR signal observed in pentacene TFT at room temperature appeared to be narrower than that observed in solution. 
The authors claimed that the narrower linewidth should 
be an evidence of a spatial extension of wavefunction.
According to CLT, the linewidth 
of a signal coming from charge distribution covering 
$N$ molecules is narrower by the factor $1/\sqrt{N}$.
Then, assuming that the signal is Gaussian 
it was concluded that the wavefunction is spread over 
$N \approx 10$ molecules. 

However, we note that the analyses were performed 
on the ESR spectrum with non-Gaussian lineshape.
Subsequent studies have shown that the field-induced
ESR signal and linewidth is temperature dependent
\cite{Matsu08,Matsu09,Maru11}. 
Typical pentacene TFTs and rubrene single-crystal transistors
exhibit sharp ESR signal whose single-Lorentzian linewidth 
presents motional narrowing \cite{KuboTo54}
effects with increase of the 
temperature. 
In case of the pentacene TFTs, the feature is well consistent 
with the thermally activated multiple trap-and-release 
(MTR) transport with the activation energy of about 
10 meV in the high temperature range. 
In contrast, the narrowing effect by the increase of 
temperature is not observed below around 50 K. 
Actually it was demonstrated by the
continuous wave saturation experiments that 
all carriers in the pentacene TFTs are localized 
at $T<50$K and all relaxation channels are frozen
at such low temperatures \cite{MMH10}. 
However the ESR spectrum is still deviated from the simple
 Gaussian at low enough temperature.
Therefore, the deviation of the ESR signal 
from Gaussian shape is not a result of relaxation
or motional narrowing but should be associated with 
the nature of weakly-localized carrier  states in the organic 
TFTs. 
It is of current important theoretical and experimental challenges 
in materials physics to understand the carrier transport of 
organic electronics devices, as they permit productions of 
large-area and flexible electronic products \cite{Bo1,Bo2}.

In the present paper we analyze the situation when
the ESR signal of a semiconducting 
organic molecular system is a smooth curve that 
deviates from the 
Gaussian linewidth even at very low temperatures
where the carriers are localized.
Here we assume that peculiar lineshape is caused by a further specific 
inhomogeneity of the pentacene TFTs, as associated with the 
distribution of weakly-localized carrier states which are 
responsible for the device operation. 
Given this assumption,  we have developed a unique technique 
for obtaining the trap density of states from a few to tens of meVs 
in pentacene TFTs \cite{MMH10} the algorithm of 
which is described in detail in this paper. 
Note that the obtained energy resolution for the trap density of 
states is much higher than that by other methods 
based on transport 
or optical measurement \cite{ad1,ad2,ad3,ad4,ad5,ad6,ad7}.
The mathematical method suggested here is rather 
general and can be applied to analyses of broad spectrum in
a variety of problems outside of those considered here.

In Section \ref{esr_tra} we study the deviation 
of the inhomogeneous low-temperature 
ESR signal from the Gaussian shape. 
We show that the signal from the one pentacene molecule 
is very similar to a Gaussian showing almost 
no individual lines from hyperfine splittings. 
It appears that the ESR lineshapes of a
signal from many independent traps of the same
kind must be a Gaussian (see \ref{singl}--\ref{gene}) 
whose width is uniquely determined by 
a single localization parameter $N_{eff}$, 
namely an effective number of sites where carrier 
is localized.
Therefore, it is concluded that the non-Gaussian lineshape 
can be ascribed to the superposition of signals
from different kinds of traps, where 
each kind of trap is described by its own 
localization parameters $N_{eff}$. 
In Section \ref{mult} we derive
an explicit relationship between the shape of the 
experimental ESR signal and the distribution of the 
traps over the localization parameters $N_{eff}$.
This relationship is a Fredholm integral equation of the first kind
where unknown function is a distribution of traps. 
Section \ref{inverse} presents an algorithm to solve 
it based on the stochastic optimization 
method (SOM) \cite{MPSS,UFN05,JPSJ06,Gunnar}. 
We describe the SOM and present an analysis of its sensitivity
to the experimental noise in Sections 
 \ref{inv_met} and \ref{tests}, respectively.
Sections \ref{exper}-\ref{practice} present  
experimental details, handling realistic noisy experimental data 
by SOM, and results 
for traps distributions in pentacene TFTs. 
Section \ref{relie} presents methods that find the limits of the 
reliability of the distributions obtained. 

Section \ref{mmh-trans} shows how
the distribution of the traps over the localization
parameter $N_{eff}$ can be mapped as a distribution 
over the binding energies $E_B$. 
This transformation can be formulated generally although 
a particular implementation of the mapping requires the 
explicit $E_B-N_{eff}$ relationship between the binding energy 
$E_B$ and the localization parameter $N_{eff}$.
This relationship for a given model 
can be obtained using the exact numeric diagrammatic Monte
Carlo method \cite{DirDMC09}, the analytic 
momentum average method \cite{BercEPL10,BercPRL09},
the coherent basis states method \cite{CBS1,CBS2},  
or numerous other methods (see [\onlinecite{FehBook}]
for a review). 
We consider in Section \ref{model-pent}
the model of  two dimensional 
Holstein polaron in the field of an on-site attractive center.
The distribution of the trapped states over the binding energies $E_B$
in pentacene TFTs are shown in Section \ref{edis-tft}. 
Sections \ref{disc} presents a discussion of our 
results and Section \ref{concl}
provides conclusive remarks.

\section{ESR spectra of trapped carriers in organic semiconductors: 
fundamental knowledge and further generalizations}
\label{esr_tra}

In this section, we introduce the well-known characteristic
 features 
of the ESR spectra of a single molecule and a 
cluster containing several molecules (\ref{singl}). 
We then present an analysis of a noticeably different 
case where the carrier is localized on a single impurity in a 
crystal (\ref{gene}).
Finally, we consider the case of traps of different 
origin where we can introduce a relationship between the 
lineshape of an experimental ESR signal and the distribution
 of the impurities over different localization parameters (\ref{mult}).

\subsection{ESR spectra for single molecule and a cluster
containing several molecules}
\label{singl}

In this section, we consider a molecular crystal in which 
single molecule contains so many nuclear spins that its ESR
spectrum has an inhomogeneous Gaussian shape.
A typical situation for a carrier trapped in a molecular 
crystal is that it is localized in a trap and its 
distribution over the molecular crystal sites 
$i$ is characterized by a probability 
distribution $\left\{ p_i \right\}$.
The temperature is assumed to be sufficiently low that we can 
neglect the ``homogeneous'' relaxation 
leading to the Lorentzian shape of the ESR signal.
It is also low enough to avoid self-averaging of the 
inhomogeneities by the "motional narrowing" mechanism.

In this case the lineshape of the ESR signal 
is determined by the ``inhomogeneous'' broadening caused by 
the site dependent distribution of the hyperfine 
interactions. 
When the typical width of the individual spectral lines of 
the split with hyperfine interaction quantum levels is larger than
the typical energy distance between these levels 
\cite{Altshuler,BookESR}, the lineshape of the ESR
signal is Gaussian. 
This shape occurs when the carrier is localized in 
either a single molecule or a cluster containing several 
molecules. 
 
The case of a carrier trapped in a crystal is noticeably 
different from the case of a cluster with several molecules. 
The probability distribution over $N$ molecules $i$ 
$\left\{ p_i, i=1,N \right\}$ is uniform $p_i=1/N$
in a cluster.
In contrast, the probability distribution in a 
crystal trap $p_i$, which is  density 
of the carrier in given site $i$,  
is not uniform, and the only 
restriction is the normalization condition 
$\sum_i p_i=1$. 
However, as shown below, the ESR signal of 
a carrier in a trap always retains the Gaussian 
shape and the width is uniquely determined by the 
carrier probability distribution $\left\{ p_i \right\}$.
 
The simplest ESR signal considered in our study is
that for a single molecule. 
The fine structure of the ESR absorption by a single molecule
in a condensed environment is frequently blurred 
by the broadening of the hyperfine levels.
The ESR signal from a single molecule in this case is Gaussian. 
The standard expression describing the hyperfine 
structure of one molecule is \cite{BookESR}
\begin{eqnarray}
R(B) & = &
\sum_{m_1=-n_1 I_1}^{n_1 I_1}
\ldots
\sum_{m_k=-n_k I_k}^{n_k I_k}
P(m_1, \dots, m_k) \times \nonumber \\
&& \frac{1}{\pi}
\frac{\Gamma}{\left(B-\sum_{i=1}^{k} A_i m_i \right)^2 + \Gamma^2} \; .
\label{one_mol}
\end{eqnarray}
Here $k$ is the number of groups of equivalent nuclei, $n_i$
is the number of equivalent nuclei in the $i$th group, 
$I_i$ is the nuclear spin in the $i$th group, $\Gamma$ is the linewidth of each peak, P is the intensity of each peak 
and $B$ is the magnetic field. 
If protons ($I=1/2$) are the only paramagnetic nuclei, as is
the case for pentacene molecules, $P$ is given as 
\begin{equation}
P(m_1, \dots, m_k)  =
{\large \Pi}_{i=1}^{k} 
\frac{C^{m_i+n_i I_i}_{2n_i I_i}}{(2I_i+1)^{-n_i}} \; ,
\label{pmmm}
\end{equation}
where $C^{m_i+n_i I_i}_{2n_i I_i}$ are binomial coefficients. 

For the particular case of the pentacene molecule 
we set $\Gamma=0.02$mT and use the coupling constants 
$\left\{ A_i; i=1, \ldots, 4 \right\}$ and numbers of equivalent
nuclei  $\left\{ n_i; i=1, \ldots, 4 \right\}$ 
reported in [\onlinecite{Bolton}]. 
\begin{figure}
        \includegraphics[scale=0.85]{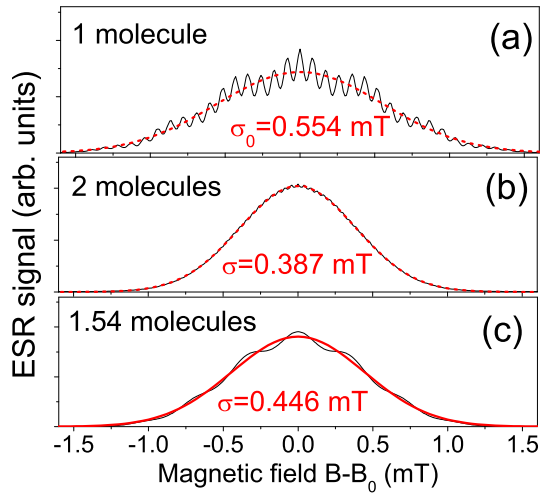}
\caption{(color online) 
Simulated ESR  spectrum (black solid lines) of 1, 2, 
and 1.54 pentacene molecules based on experimental 
hyperfine splitting of one pentacene molecule with
fits by Gauss distribution (red dashed lines). 
}
\label{fig1}
\end{figure}
The ESR signal obtained from Eqs. (\ref{one_mol})
and (\ref{pmmm}) can be represented (see Fig.~\ref{fig1}a) as
a curve fluctuating around the Gaussian envelope
\begin{equation} 
G_0(B)=\sqrt{\frac{1}{2 \pi \sigma_0^2}}
\exp \left[ -  
\frac{(B-B_0)^2}{2 \sigma_0^2}
\right]
\label{G0}
\end{equation}
with standard deviation $\sigma_0 = 0.554$ mT.

The standard situation, known from the physics 
of gases and solutions, is the case where the carrier 
is localized in a cluster containing $N$ molecules and its
density is spread over $N$ molecules.
In this case the signal retains its Gaussian shape with 
the width of the distribution reduced by the factor $N^{1/2}$. 
The hyperfine structure of the $N$ molecules 
can be simulated by Eqs.  
(\ref{one_mol}) and (\ref{pmmm}) by replacing 
$n_i \to N n_i$ and $A_i \to A_i / N$.
Figure~\ref{fig1}b shows an example of the spectrum 
for $N=2$ with standard deviation  
$\sigma=\sigma_0/\sqrt{2}$.  
It is clear that the oscillations around the 
Gaussian envelope are quickly suppressed as
$N$ increases. 

The shape and the $1/\sqrt{N}$ narrowing
factor of the ESR signal for a carrier distributed 
over a cluster with $N$ molecules follow from the CLT. 
The shifts of the signal, $y_i=(B_i-B_0)$ for 
the $i$-th molecule, is an independent random variable   
with Gaussian distribution $R(y)=G_0(y)$ with 
standard deviation $\sigma_0$. 
By the CLT, the distribution $R(\bar{y})$ 
of the random variable 
$\bar{y} = N^{-1} \sum_{i=1}^{N} y_i$ is also Gaussian 
with $\sigma_N = \sigma_0 / \sqrt{N}$.

\subsection{ESR spectra for trap in crystal}
\label{gene}

\begin{figure}
        \includegraphics[scale=0.99]{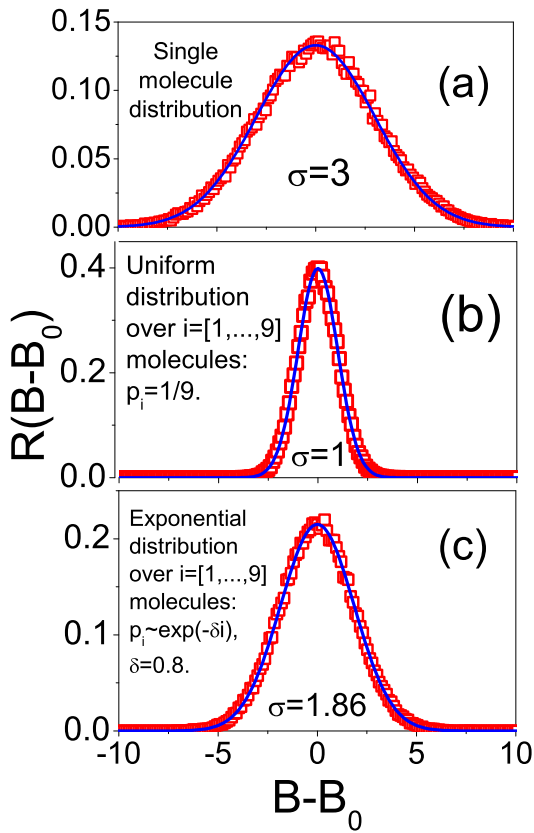}
\caption{(color online) 
(a) Generated Gauss distribution with $\sigma=3$,
(b) distribution of a uniformly weighted $p_i=1/9$ sum 
of 9 normally distributed random variables, and (c) 
distribution of exponentially weighted 
$p_i=\exp(-\delta i)$ ($\delta=0.8$) sum of 9 normally 
distributed random variables. The squares show the 
distribution of the generated random variables and 
the solid lines are the fits by Gaussian functions.
The envelope Gaussian function is normalized to unity in all
panels.}
\label{fig2}
\end{figure}

The situation for a carrier localized in a trap
in a crystal is different from the above situation
with $N$ molecules.
The latter case assumes uniform charge 
distribution, and thus the CLT applies.    
In contrast, the distribution over molecules $i$  
in a trap $\left\{ p_i \right\}$ is nonuniform with the  
probabilities $p_i$ subjected to the normalization condition 
$\sum_i p_i \equiv 1$. 
Hence, we cannot assume a Gaussian lineshape
for the ESR signal; the lineshape must be studied separately.

Regardless of the lineshape, the probability
distribution $\left\{ p_i \right\}$ unambiguously 
determines the linewidth of the ESR signal.
The linewidth is characterized by the 
standard deviation $\sigma$ which is the 
root square of the second moment of the 
linewidth.  
If we consider the standard deviation of a signal in 
a trap $\sigma$ and compare it with that from 
a single molecule $\sigma_0$ we can introduce 
the effective number of molecules 
$N_{eff}(\{ p_i \})$ to describe the linewidth
of the ESR signal from a carrier in a trap.   
The distribution of the ESR shift $B$ 
is the same for each molecule $i$ with mean 
$\left\langle B \right\rangle=B_0$ and 
variance 
$\sigma_0^2 = \left\langle (B-B_0)^2 \right\rangle$.
Since the hyperfine configuration of molecules are 
independent of each other the variables $y_i=(B_i-B_0)$
are independent for different molecules $i$.
Hence,  the standard deviation 
$\sigma(\left\{ p_i \right\})$ of the sum of random 
variables $\bar{y}=\sum_i p_i (B_i-B_0)$ is related to the
single-molecule standard deviation $\sigma_0$ by 
the expression
\begin{equation} 
\frac{\sigma( \left\{ p_i \right\} )}{\sigma_0} 
= \sqrt { \sum_i p_i^2 }  \;.
\label{sigma} 
\end{equation}
It is natural to define the effective
number of molecules $N_{eff}(\{ p_i \})$, 
corresponding to the charge distribution $\{ p_i \}$, to be 
\begin{equation}
\sigma(\{ p_i \}) = \sigma_0 / \sqrt{N_{eff}(\{ p_i \})} \; .
\label{Ndef}
\end{equation}
Then, the effective number of molecules 
$N_{eff}(\{ p_i \})$ is unambiguously determined by the 
charge distribution in a trap $\{ p_i \}$
\begin{equation}
N_{eff}(\{ p_i \}) = \left[ \sum_i p_i^2 \right]^{-1} \; . 
\label{Neff}
\end{equation} 

To study the shape of the ESR signal for a trap
with charge density $\{ p_i \}$ we generated 
by a standard method \cite{MPSS} random variables 
$\{ y_i \}$ following a Gaussian distribution
with dispersion $\sigma_0=3$ (Fig.~\ref{fig2}a).
Then, we studied the distribution of the
random variable $\bar{y}=\sum_i p_i y_i$.
By the CLT the uniform distribution $p_i=1/N$ 
leads to a Gaussian shape of the signal with the dispersion 
narrowed by the factor $\sqrt{N}$ (Fig.~\ref{fig2}b). 
After performing simulations with a large set of 
different distributions $p_i$ we conclude 
that the distribution of the random variable 
$\bar{y}=\sum_i p_i y_i$ is always   
Gaussian (see Fig.~\ref{fig2}c)
with $N_{eff}(\{ p_i \})$ defined 
by Eq.~(\ref{Neff}).

Hence, we conclude that the shape of the ESR signal
for carriers localized in a set of identical 
{\it independent} traps is uniquely determined by the distribution 
density  $\left\{ p_i \right\}$.
It is always Gaussian with the standard deviation
$\sigma$ defined by Eqs.~(\ref{Ndef}) and (\ref{Neff}).

\subsection{ESR spectra for several kinds of traps}
\label{mult}

Since the ESR signal for independent 
identical traps is always Gaussian, we 
assume that a non Gaussian shape originates 
from the superposition of the signals 
from different kinds of traps.
Indeed, each different kind of trap is characterized 
by a different probability distribution of the trapped 
carriers.

To describe the experimental spectrum ${\cal E}_{exp}(B)$
by the superposition of the ESR spectra 
${\cal G}(B,\xi)$ for each trap type $\xi$ 
we must choose a parameter $\xi$ that unambiguously 
characterizes the spectrum ${\cal G}(B,\xi)$.
It follows from the analysis in Section \ref{singl}
that the ESR spectrum from identical traps is Gaussian and
can be characterized by a single parameter $N_{eff}$.
This parameter reflects the spatial distribution 
$\left\{ p_i \right\}$ of a charge in a trap (\ref{Neff}) and 
determines the narrowing of the Gaussian ${\cal G}(B,\xi=N_{eff})$ width 
$\sigma(\{ p_i \}) = \sigma_0 / \sqrt{N_{eff}(\{ p_i \})}$ with respect to the ESR width $\sigma_0$ of a carrier localized on a single molecule.
Therefore, the ESR signal for the same trap type,
characterized by the same spatial extension $\xi=N_{eff}$, 
can be expressed as  
\begin{equation} 
G(B,N_{eff})=\sqrt{\frac{N_{eff}}{2 \pi \sigma_0^2}}
\exp \left[ -
\frac{(B-B_0)^2}{2 (\sigma_0^2/N_{eff})}
\right] \; .
\label{GN}
\end{equation}

Introducing the distribution function $D(N_{eff})$
of the traps in pentacene over $N_{eff}$ 
we can express the experimental signal 
${\cal E}_{exp}(B)$ in terms of the superposition  
\begin{equation} 
{\cal E}_{exp}(B) = \int_{1}^{\infty} G(B,N_{eff}) D(N_{eff}) d N_{eff} 
\label{conv1}
\end{equation}
which is a convolution of the distribution function 
$D(N_{eff})$ of the traps and the Gaussian signal for a 
trap type characterized by the localization
parameter $N_{eff}$. 

Most techniques for ESR measurements detect
the derivative ${\cal E}_{exp}(B)$ over the magnetic 
field $B$ and hence the experimental signal 
${\cal X}_{exp}(B) = d {\cal E}_{exp}(B)/dB$ 
is related to the distribution function of the traps 
$D(N_{eff})$ via 
\begin{equation} 
{\cal X}_{exp}(B) = \int_{1}^{\infty} 
\frac{dG(B,N_{eff})}{dB} D(N_{eff}) d N_{eff} \; .
\label{conv2}
\end{equation}

Hence, to obtain the distribution $D(N_{eff})$
we must solve one of the integral equations 
(\ref{conv1}) and (\ref{conv2}).
The experimental signals ${\cal X}_{exp}(B)$ 
(${\cal E}_{exp}(B)$)
and the kernel $dG(B,N_{eff})/dB$ ($G(B,N_{eff})$)
are known functions and the distribution 
$D(N_{eff})$ is to be determined.

\section{From ESR spectrum to trap distribution:
conversion of experimental lineshape into 
distribution of traps over degree of localization}
\label{inverse}

Equation (\ref{conv1}) [(\ref{conv2})], where 
${\cal E}_{exp}(B)$ [${\cal X}_{exp}(B)$] 
and $G(B,N_{eff})$ [$dG(B,N_{eff})/dB$] are known and 
$D(N_{eff})$ is to be determined, is a Fredholm equation 
of the first kind. 
Naively, to find the solution $\widetilde{D}(N_{eff})$  
we must maximize the inverse deviation
\begin{equation} 
Q = \left\{ \int_{B_{min}}^{B_{max}} 
\left| 
{\cal X}_{exp}(B) - \widetilde{\cal X}(B) 
\right| \right\}^{-1} \; .
\label{measu}
\end{equation}
Here $B_{min}$ ($B_{max}$) is the lower (upper) bound
of a magnetic field where the signal is larger
than the experimental noise.
${\cal X}_{exp}(B)$ are experimental data and  
$ \widetilde{\cal X}(B)$ is obtained from 
the distribution $\widetilde{D}(N_{eff})$ 
\begin{equation}
\widetilde{\cal X}(B) =\int_{0}^{\infty} 
\frac{dG(B,N_{eff})}{dB} \widetilde{D}(N_{eff}) d N_{eff} \; 
\label{tilx}
\end{equation}
which is considered to be a solution of the integral equation.
However, such naive approach leads to huge unrealistic 
oscillations of the solution $\widetilde{D}(N_{eff})$.  
Instead, we need to apply one of the advanced techniques
developed for such equations. 

Section \ref{inv_met} gives a general description
of the stochastic optimization method (SOM) 
\cite{MPSS,UFN05,JPSJ06,Gunnar} 
for the solutoion of Eqs. (\ref{conv1}) and (\ref{conv2}). 
Section \ref{tests} applies the 
method to the analysis of the ESR data
and demonstrates the influence of experimental noise on
the reliability of the results.
Section \ref{exper} presents an 
experimental technique to obtain ESR spectra 
suitable for the fine analysis of the data.
Section \ref{practice} introduces an algorithm that implements 
the SOM and presents results for the trap distribution in
pentacene TFTs. 
Finally, Section \ref{relie} demonstrates 
the limits of the reliability of the distribution obtained by
solving Eqs. (\ref{conv1}) and (\ref{conv2}).

\subsection{Method to solve inverse problem}
\label{inv_met}

It is notoriously difficult to solve the Eqs.
 (\ref{conv1}) and (\ref{conv2}) because 
these equations belong to the class of "ill posed" 
problems. 
Naively, the true solution of the Eq.  (\ref{conv2})
$\widetilde{D}(N_{eff})$, being convoluted with the kernel,
produces the function $\widetilde{\cal X}(B)$, -
which coincides with the given function ${\cal X}_{exp}(B)$.
However, the general feature of the practical implementations
of Eq. (\ref{conv2}) is that the knowledge about the function 
${\cal X}_{exp}(B)$ is ``noisy'' and incomplete. 
Specifically, ${\cal X}_{exp}(B)$ is known for a particular set 
of points $\{B_i, i=1,M\}$ with some errorbars resulting from
the experimental noise.
In this case, to find a "solution", 
we can introduce the residual function 
\begin{equation}
\Delta(i) = \widetilde{\cal X}(B_i)- {\cal X}_{exp}(B_i), \;\; i=1,M 
\label{residual}
\end{equation}     
and optimize a measure of the deviation of 
"the solution" $\widetilde{\cal X}(B)$ from the given 
data ${\cal X}_{exp}(B)$.  
For example, we could maximize the inverse deviation (\ref{measu})
\begin{equation}
Q=\left[ \sum_{i=1}^M \left| \Delta(i) \right| \right]^{-1} \; .
\label{ap4}
\end{equation}
Naturally, $\Delta(i)$ is never equal to zero 
at all points $i=1,M$ when realistic noisy 
data $\{{\cal X}_{exp}(B_i), i=1,M\}$ are considered.
Hence, even the best measure  
$Q^{max}$ is not equal to infinity. Therefore,
the only feasible strategy for Eq. (\ref{conv2}) is to find a solution, 
that is "the best'' in some sense.

The above features are the characteristics of the class
of "ill posed" problems for which we can not get an  
exact solution and can only find the "best" choice of 
$\widetilde{D}(N_{eff})$ for the given data set 
$\{{\cal X}_{exp}(B_i), i=1,M\}$.    
The naive approach, where we simply maximize measure 
(\ref{ap4}), leads to unreasonable "solutions".
Typically, they have huge fluctuations which exceed
the true values of $D(N_{eff})$ by several 
orders of magnitude.  
To get a reasonable description of $D(N_{eff})$ 
we must suppress this "saw tooth" 
instability. 

There are two different strategies. 
The first is the regularization 
method, e.g. the popular maximal entropy method 
\cite{MaxEnt} as one of many such methods \cite{Tikhonov,Perchik}. 
This approach maximizes a measure that is similar to 
(\ref{ap4}) but modified in such a way that the solution is 
smooth enough to suppress "saw tooth noise".
The main drawback is that the solution is corrupted by the 
smoothing regularization procedure. 
The second strategy uses modern stochastic approaches 
to obtain many statistically independent solutions 
(see \cite{MPSS,Gunnar} and the references therein)
whose linear combination smoothes the "saw tooth noise"
without corrupting individual solutions. 
Since this approach has been shown \cite{Gunnar}
to be better for the solution of Eq. (\ref{conv2}), 
we applied the SOM \cite{MPSS,UFN05,JPSJ06} as a particular
example of the stochastic technique \cite{Gunnar}.

\begin{figure}
        \includegraphics[scale=0.36]{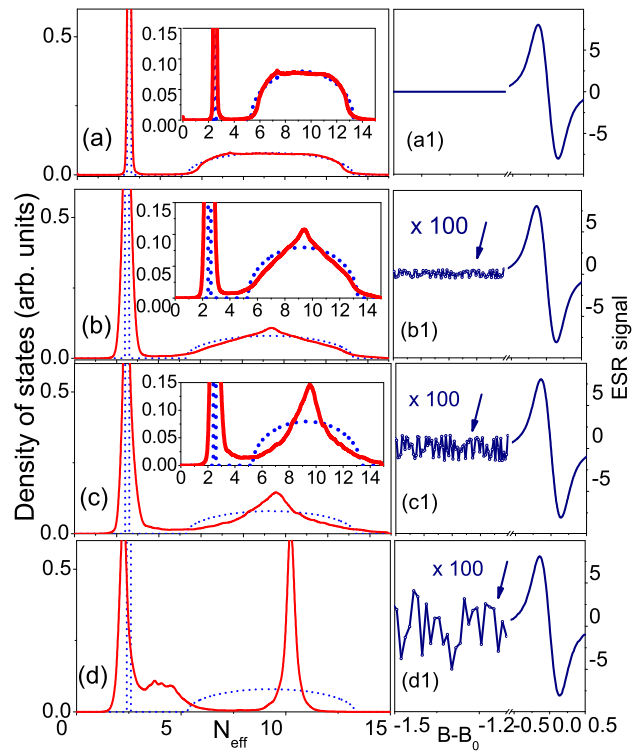}
\caption{(color online) 
Test of SOM procedure to restore distribution 
of spatial extent of traps $D(N_{eff})$
from ``experimental'' ESR data ${\cal X}_{exp}(B)$ 
(\ref{conv2}) under different levels of noise.
``Experimental'' ESR data with noise 
are given in panels (a1)-(d1) and actual (restored)
spectrum  $D(x)$ is indicated by dashed (solid) line in
corresponding panels (a)-(d). 
Noise level is (a,a1) $f=0.0$, (b,b1) $f=0.01$
[$s_n=10^{-3}$], (c,c1) $f=0.03$ [$s_n=3*10^{-3}$], 
and (d1,d) $f=0.1$ [$s_n=10^{-2}$].  
Restored spectrum is obtained by solving  
Eq. (\ref{conv2}) using SOM. 
Insets show details of corresponding spectra.
}
\label{fig5_1}
\end{figure}
\begin{figure}
        \includegraphics[scale=0.36]{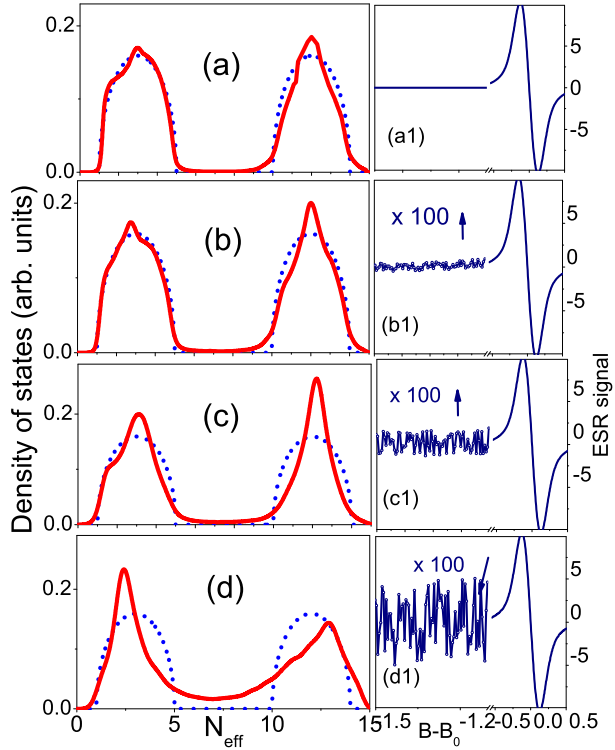}
\caption{(color online) 
See figure caption in Fig. \ref{fig5_1}.}
\label{fig5_2}
\end{figure}

The function $D(N_{eff})$ is a distribution function 
and, hence, it is non negative $D(N_{eff}) \ge 0$ and normalized   
\begin{equation}
\int_a^b D(N_{eff}) dN_{eff} = I 
\label{ap2}
\end{equation}
to a certain number $I$. 
The noise present in experimental data 
$\{{\cal X}_{exp}(B_i), i=1,M\}$ 
makes the problem of normalization nontrivial. 
Some approaches to handle the normalization 
problem are suggested in the next chapters.

\subsection{Tests of SOM and 
role of noise in experimental data}
\label{tests}

The SOM has been successfully applied to many integral
equations.  
The kernels of the equations are different 
from those in Eqs. (\ref{conv1}) and (\ref{conv2}).
The exponential kernel $K(y,x)=\exp[-yx]$ was  
examined in
[\onlinecite{MPSS,s01,s02,s03,s04,s05,s06,s07,s08,s09,s10,s11,s12,s13,s14,s15,s16,s17,s18,s19}],
and various kernels ranging from the Fermi distribution
to the Matsubara frequency representation are considered in 
[\onlinecite{Licht1,Licht2,Licht3,Licht4}].
We test the applicability of the SOM to the kernels
of Eqs. (\ref{conv1}) and (\ref{conv2}).  
We also show how the statistical noise of the 
experimental data can obscure the information 
that can otherwise be obtained by solving
these equations.

To verify that the SOM is applicable to Eqs
 (\ref{conv1}) and (\ref{conv2}) with the kernel 
defined by (\ref{GN}) we introduced a 
normalized to unity function $D(N_{eff})$ (see the dotted line in 
Figs.~\ref{fig5_1} and \ref{fig5_2}) and 
generated a set of "experimental'' data 
$\{{\cal X}_{exp}(B_i), i=1,200\}$ 
using relations (\ref{conv1},\ref{conv2}). 
Then, we attempted to find 
$\widetilde{D}(N_{eff})$ by solving 
Eqs. (\ref{conv1}) and (\ref{conv2}).
For ideal data without noise 
in the set $\{{\cal X}_{exp}(B_i), i=1,200\}$ we were
able to restore $D(N_{eff})$ successfully
(see Fig.~\ref{fig5_1}(a) and \ref{fig5_2}(a)). 
We did not find a significant difference between the 
results obtained by solving Eqs. (\ref{conv1}) and (\ref{conv2}). 

In spite of the extensive usage of SOM \cite{MPSS,s01,s02,s03,s04,s05,s06,s07,s08,s09,s10,s11,s12,s13,s14,s15,s16,s17,s18,s19,Licht1,Licht2,Licht3,Licht4} there is no precise understanding of how the errorbars of the 
data $\{{\cal X}_{exp}(B_i), i=1,M\}$ obscure the solution 
$\widetilde{D}(N_{eff})$. 
The difficulty is the dependence of the result 
$\widetilde{D}(N_{eff})$ on the level of statistical noise, 
the shape of $D(N_{eff})$, the kernel, and even 
the number of points $M$ in the data
set $\{{\cal X}_{exp}(B_i), i=1,M\}$.
Therefore, below we present some examples
showing how the noise of the experimental data 
obscures the solution of Eq. (\ref{conv2}). 
These {\it ad hoc} examples indicate the 
general trends but must not be treated as a quantitative 
analysis of the influence of noise on the reliability of 
the SOM results.
These data should not be used for the analysis of other 
cases.
In Section \ref{relie} we present a procedure to check the
reliability of particular solutions.  
     
To introduce noise we used a sequence of 
random numbers ${\cal R}(i)$  uniformly
distributed in the range [-1,1] and 
generated data sets   
$\{{\cal X}_{exp}(B_i)+ (f/2) {\cal R}(i), i=1,200\}$,
referred in the following as "experimental"
ESR data, 
with different amplitudes of the noise $f$.
The signal-to-noise ratio is defined
as a ratio of the amplitude $f$ and 
maximal absolute value $MAX\{ \mid {\cal X}_{exp}(B_i) \mid \}$ 
of the given signal ${\cal X}_{exp}(B_i)$:  
as $s_n = f / MAX\{\mid {\cal X}_{exp}(B_i) \mid$.
The results presented in Figs~\ref{fig5_1} and 
\ref{fig5_2} illustrate the general 
trends. 
An increase in the signal-to-noise ratio $s_n$ corrupts 
the solution for large values of $N_{eff}$ first:
the shape of the high-energy peak is not reproduced 
but its position is still correct.
At higher values of $s_n$ the shape of the 
low-energy peak is not reproduced.  
A comparison of Figs~\ref{fig5_1} and 
\ref{fig5_2} shows that the spectrum with 
a sharp feature at small $N_{eff}$ (Fig.~\ref{fig5_1})
is more robust to experimental noise than 
that with a broad feature at small $N_{eff}$
(Fig.~\ref{fig5_2}).  
Note that although the shapes of the high- and
low-energy peaks are not reproduced, their 
positions are still approximately 
correct even for large values of the signal-to-noise 
ratio $s_n$.

\subsection{Experimental data for analysis}
\label{exper}
 
To conduct reliable spectral analysis as discussed
above, we need high-precision ESR spectra for the carriers in
organic TFTs. We acquired the spectra by the
following procedures: We used a commercially-available
X-band (9 GHz) ESR apparatus (JES-FA200, JEOL) equipped with a
high Q cylindrical cavity (Q factor 4000-6000 for the TE$_{011}$
mode). We fabricated bottom-gate, top-contact pentacene TFTs
with high mobility that are suitable for 
high-precision measurements. The device is composed of a
100-$\mu$m-thick poly(ethylene naphthalate) (PEN) film as a
nonmagnetic substrate, a 1-$\mu$m-thick Parylene C 
film as a gate dielectric layer (4.5 nF/cm$^2$), and a 
50-nm-thick pentacene film as
the semiconducting layer. The gate, source, and drain electrodes
are composed of vacuum-deposited gold films with a thickness of
30 nm; this is much thinner than the skin depth of gold (about
790 nm).

Since the field-induced carrier is accumulated only at the
semiconductor/insulator interface, the ESR signal is
proportional to the total channel area of the TFTs. We used
a device with a width of 2.5 mm and a length of 20 mm, 
the dimension of which is limited by the inner
diameter of the ESR tube and
the cavity size. We used a stack of ten sheets of TFTs
for the high-precision ESR measurement, to
obtain field-induced carriers ten times as large as those in a
single sheet. The total carrier number at $V_G$ = -200V is
estimated as 2.8x10$^{13}$.

The semiconducting pentacene layer is composed of a
uniaxially-oriented polycrystalline film where all the component
pentacene molecules aligned with the molecular long axes are roughly
perpendicular to the film plane. In the measurement, a static
magnetic field was set perpendicular to the film plane to
eliminate the anisotropic effect of the g tensor.
A continuous-flow cryostat was used for the low-temperature
measurements. In the FESR measurements at low temperature, we
first applied the gate voltages at room temperature 
(with the source and
drain shorted) and then cooled the device to the set
temperature, to avoid a delay in the charge
accumulation. The temperature was stabilized carefully so that
the fluctuation at 20 K was about 0.01 K, which minimized the
effect of temperature-dependent spin susceptibility.

\subsection{Practical implementation of method: 
Distribution of traps in pentacene TFT}
\label{practice}

Equation (\ref{conv2}) is preferable for the practical 
implication of the SOM. 
The problem with realistic noisy data from ESR
experiments is that there is some uncertainty in the normalization 
and background of the experimental data. 
In the ideal case, implying the normalization of the 
distribution $D(N_{eff})$ to unity and the conditions 
$\int_{-\infty}^{\infty} G(B,N_{eff})d N_{eff}=1$ and 
$\int_{-\infty}^{\infty}dN_{eff} \int_{-\infty}^{B} dz [dG(z,N_{eff})/dz]=1$, 
we must normalize the experimental data as 
\begin{equation}
\int_{-\infty}^{\infty}dB {\cal E}_{exp}(B) = 1
\label{norm1}
\end{equation}
and 
\begin{equation}
\int_{-\infty}^{\infty}dB  \int_{-\infty}^{B}dz {\cal X}_{exp} (z) = 1 \; ,
\label{norm2}
\end{equation}
respectively. 
However, the normalization of the experimental data is not 
exact because experimental noise can lead to uncertainty 
of a few percentage points.
On the other hand, the solution of Eq. (\ref{conv2}) is 
sensitive to possible normalization errors. 

There is also a problem with the background. 
If the experimental data are obtained in the form ${\cal E}_{exp}(B)$ 
(see e.g. Fig.~\ref{fig2}), there is no unique procedure
to determine the constant background level that must 
be subtracted from the data to get a pure signal 
for the ESR transition. 
This uncertainty also increases the uncertainty relating to the 
normalization of the data.  
However, the problem of the unknown background 
disappears when Eq. (\ref{conv2}) is used as the
integral equation because  
$\lim_{B \to \pm \infty}{\cal X}_{exp} (B) = 0$
for any constant background. 
Therefore, to analyze the ESR data for pentacene TFTs
we used (\ref{conv2}) where the only uncertainty is that 
of normalization. 
\begin{figure}
        \includegraphics[scale=0.8]{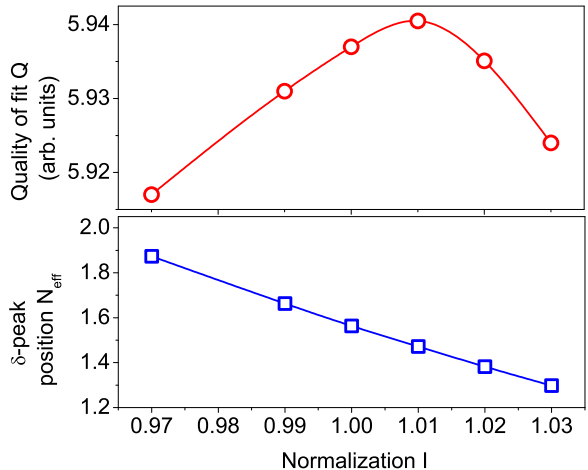}
\caption{(color online) 
(a) Dependence of the fit quality $Q^{max}$ and position 
of low-energy peak $\delta(N-N_{eff})$ on normalization
$I$ of the spectral function.
}
\label{fig6}
\end{figure}

\begin{figure}
        \includegraphics[scale=1.2]{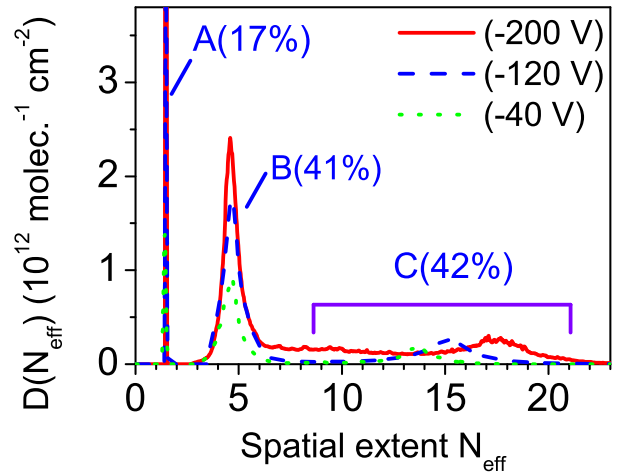}
\caption{(color online) 
Distribution of trap states in pentacene TFT versus
spatial extent $N_{eff}$ of charge 
distribution in traps obtained
from ESR spectrum of pentacene TFT at 20 K
with gate voltage -200 V (solid line), 
-120 V (dashed line), and -40 V (dotted line).
}
\label{fig4}
\end{figure}

To handle the normalization uncertainty
we can change the normalization in either 
Eq.~(\ref{ap2}) or Eq.~(\ref{norm2}). 
The two choices are equivalent because of the linearity
of Eq. (\ref{conv2}). 
In practice we normalized the experimental 
signal as shown in (\ref{norm2}) and varied the 
normalization $I$ of the distribution density 
$\widetilde{D}(N_{eff})$ (\ref{ap2}). 
We handled the normalization uncertainty for the result  
shown in Fig. \ref{fig4} as follows.
The spectrum at gate voltage -200 $V$ was considered
(the result for this gate voltage is 
shown in Fig.~\ref{fig4} by solid line).  
The integral equation (\ref{conv2}) was solved for 
different normalizations $I$ and the solution with the 
normalization having the best deviation 
measure $Q^{max}$ was chosen.  

Figure~\ref{fig6} shows the best inverse deviation 
$Q^{max}$ (\ref{measu},\ref{ap4})
and the position of the sharp peak in $D(N_{eff})$ 
versus the normalization $I$.
It can be seen that the position of the sharp peak is  
sensitive to the value of the normalization $I$ and 
this may be a source of the volatility in 
the fine analysis of the ESR data. 
However, it can be shown that the suggested approach 
to the the normalization $I$, which leads to the 
best deviation measure $Q^{max}$, is robust 
and produces stable results.
We have demonstrated this via analysis of the ESR data at different 
gate voltages (Fig.~\ref{fig4}).  
We found that the best normalizations 
are different at different gate voltages:
$I(V=-200)=1.01$, $I(V=-120)=1.038$, and 
$I(V=-40)=1.03$, respectively. 
However, the position of the sharp peak at low 
values of $N_{eff}$ does not depend on the voltage $V$ 
if at each voltage $V$ we use the normalization $I(V)$ 
that corresponds to the best deviation measures $Q^{max}$. 
Physically, the sharp peak at low values of $N_{eff}$ corresponds to 
deep impurity levels that depend only slightly on the gate voltage. 
Therefore, its independence on the gate voltage in the fine 
analysis of the ESR spectra indicates the high stability of 
the procedure based on the suggested approach.

\subsection{Reliability of trap distribution result}
\label{relie}

Since solving Eq. (\ref{conv2}) is an ''ill-posed'' problem
it is useful to understand how much information we can get 
from the analysis and to check how many details of the 
resulting distribution of impurities are reliable.  
The reliability can be analyzed by plotting the 
residual function (\ref{residual}).

\begin{figure}
        \includegraphics[scale=0.99]{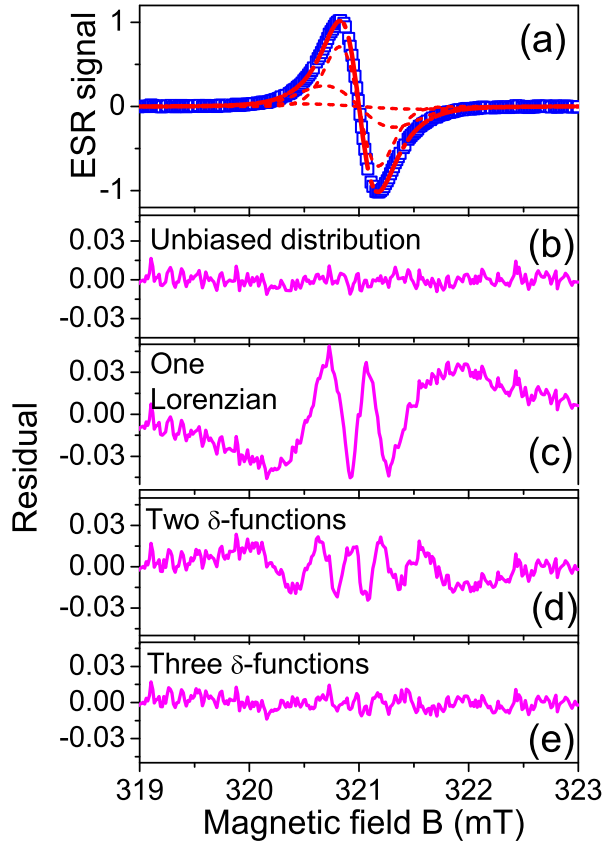}
\caption{(color online)
(a) Experimental signal (squares), fit by spectrum in
Fig.~\ref{fig4} at the gate voltage -200 $V$ 
(solid lines) and contributions from 
(see Fig.~\ref{fig4}) A, B, and C components (dashed lines). 
Residuals  (\ref{residual})
${\cal X}_{exp}(B)-\widetilde{\cal X}(B)$
from (b) unbiased distribution (Fig.~\ref{fig4}), 
(c) best fit by one Lorentzian, (d) best fit by two 
$\delta$-functions $0.68\delta(N-4.5)$ and 
$0.32\delta(N-20)$, and (e) best fit by three $\delta$-functions 
$0.31\delta(N-1.4)$, $0.51\delta(N-7.5)$, and 
$0.18\delta(N-25.0)$.        
}
\label{fig7}
\end{figure}

The spectrum $\widetilde{D}(N_{eff})$ (Fig.~\ref{fig4}) obtained 
by solving Eq. (\ref{conv2}) has three peaks. 
Figure.~\ref{fig7}a shows the fit of the ESR signal using the 
distribution $\widetilde{D}(N_{eff})$ in Fig.~\ref{fig4}.  
It also shows the separate contributions of the A, B, 
and C components of the distribution.
To clarify which features of the distribution $\widetilde{D}(N_{eff})$ 
are reliable for the given level of noise in the 
experimental data we studied the residuals (\ref{residual}) 
${\cal X}_{exp}(B)-\widetilde{\cal X}(B)$ 
(Figs.~\ref{fig7}b-e). 
We can see that the quality of the fit by the SOM 
(Fig.~\ref{fig7}b) is much better than that obtained by
e.g., the Lorentzian (Fig.~\ref{fig7}c) and two $\delta$-functions
(Fig.~\ref{fig7}d). 
The fit from three $\delta$-functions gives a residual function
(Fig.~\ref{fig7}e) as good as that
obtained from $\widetilde{D}(N_{eff})$ in Fig.~\ref{fig4}. 
Therefore, we conclude that, within the limits of the 
noise of the experimental data, the existence of at least 
three kinds of traps is a reliable result.    

We note that the distribution over the parameter 
$N_{eff}$ in Fig.~\ref{fig4} is free from any assumption 
about the shape of the distribution.
Indeed, because of the noise of the current experimental 
data, the only reliable conclusion is the statement about the 
existence of at least three types of traps. 
However, a data analysis with less noise could, in principle, 
reveal more fine structure in the distribution 
function $D(N_{eff})$.

\section{Transformation from spatial distribution 
to energy distribution}
\label{mmh-trans}

In this section we discuss finding the distribution
of the traps ${\cal Z}(E_B)$ over the binding energies $E_B$ given
the distribution $D(N_{eff})$ over the localization parameter
$N_{eff}$. 
This transformation is trivial when there is {\it a priori} 
knowledge of the functional dependence 
\begin{equation}
N_{eff}=N_{eff}(E_B) \; .
\label{nvse}
\end{equation} 
Indeed, there is a balance relation 
\begin{eqnarray}
&{\cal Z}\left( \frac{E_B^{(i+1)}+E_B^{(i)}}{2} \right)  
\left[ E_B^{(i+1)}-E_B^{(i)} \right] & = \nonumber \\ 
&D\left( 
\frac{N_{eff}\left(E_B^{(i+1)}\right)+N_{eff}\left(E_B^{(i)}\right)}{2} 
\right) & \nonumber \\ 
&\left[ N_{eff}\left(E_B^{(i+1)}\right)-N_{eff}\left(E_B^{(i)}\right) \right] &
\nonumber
\label{bal}
\end{eqnarray} 
between $D$ and ${\cal Z}$ for two nearby points $E_B^{(i+1)}$ and $E_B^{(i)}$.
Then, in the limit  $E_B^{(i+1)} \to E_B^{(i)}$ we get   
\begin{equation}
{\cal Z}(E_B) = D(N_{eff}) \frac{d N_{eff}(E_B)}{d E_B} \; .
\label{transf}
\end{equation}
The above generic relation (\ref{nvse}) can be obtained as a 
parametric function provided we can 
to choose an appropriate model and calculate  
$N_{eff}=N_{eff}(\{ {\cal P} \})$ and $E_B=E_B(\{ {\cal P} \})$ 
in some domain $\{ {\cal P} \}$ of the parameters ${\cal P}$. 

The general strategy for the transformation $D \to {\cal Z}$ 
does not depend on the model.
However, the specific form of relation (\ref{nvse}) 
depends on the model chosen to describe the localization 
of a carrier in a trap.
To study carrier trapping in pentacene TFTs 
we consider a model of a two-dimensional (2D) Holstein 
polaron in a field of an attractive center.   
This model is considered in Section \ref{model-pent}, and the 
results of the transformation for pentacene TFTs are given in 
Section \ref{edis-tft}. 

\subsection{Model for traps: 2D Holstein polaron in field of 
on-site attractive center}
\label{model-pent}

It is already well known that the behavior of a carrier in 
pentacene TFTs can be described by a particle in 
a system with attractive impurities \cite{Matsu08}. 
It is also known that the carrier is subject to the electron-phonon 
interaction \cite{Bao07}. 
To model the behavior of a carrier in pentacene TFTs we chose 
the simplest Hamiltonian describing a 2D Holstein polaron in a 
field of on-site attractive center
\begin{eqnarray}
&H =
-t\sum_{\langle {\bf i},{\bf j} \rangle} c_{\bf i}^{\dagger} c_{\bf j} 
+\omega_{\mbox{\scriptsize ph}} \sum_{\bf i} b_{\bf i}^{\dagger} b_{\bf i}&
\nonumber \\
&- \gamma \sum_i (b_{\bf i}^{\dagger}+b_{\bf i})  
c_{\bf i}^{\dagger} c_{\bf i} 
- U c_{\bf 0}^{\dagger} c_{\bf 0}& \; .
\label{H}
\end{eqnarray}
\begin{figure}
        \includegraphics[scale=0.99]{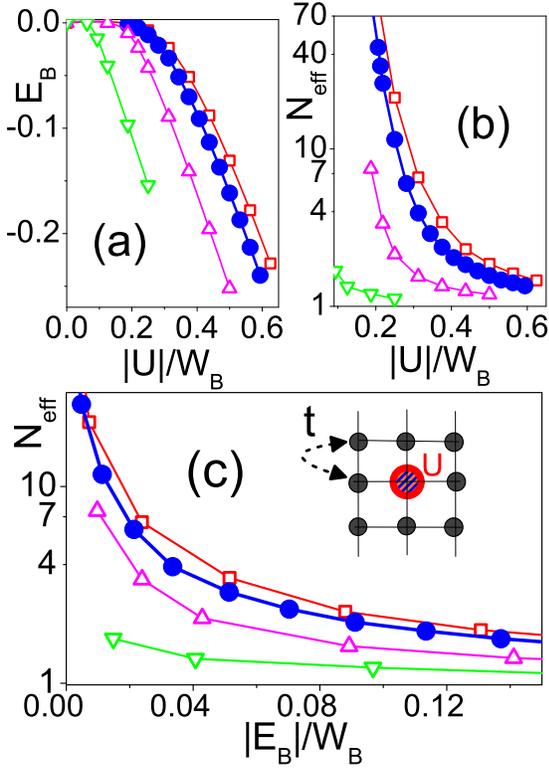}
\caption{(color online) 
Dependence of (a) binding energy $E_B$ in units of 
bandwidth $W_B=8t$ and (b) effective number of molecules 
$N_{eff}$ on absolute value of potential of attractive 
impurity $\mid U \mid$ in units of $W_B$. 
(c) Dependence of effective localization number $N_{eff}$ 
on binding energy $E_B$ in units of 
bandwidth $W_B$. Curves are presented for $\lambda=0$ (squares),
$\lambda=0.15$ (circles), $\lambda=0.5$ (triangles pointing up), and
$\lambda=1$ (triangles pointing down). Lines are to guide the eye.     
Inset in (c) shows schematically model represented by 
Hamiltonian (\ref{H}). 
}
\label{fig8}
\end{figure}
Here, $c^\dagger_i$ ($b^\dagger_i$) is the creation operator 
for the carrier (phonon) in the  i-th molecule.  
$U$ is the attractive impurity potential for the 
carrier $c^\dagger_0$ at  site $0$ and 
$\omega_{\mbox{\scriptsize ph}}$ is the frequency of the dispersionless phonon. 
The amplitude $t$ describes the 
electron transfer $\propto t$ between nearest neighbor sites 
and the local Holstein coupling to the phonons is $\propto \gamma$.
\begin{figure}
        \includegraphics[scale=0.8]{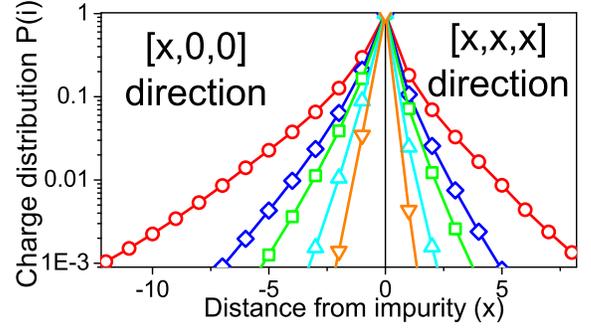}
\caption{(color online) 
Charge distributions around attractive impurity
with potential U at $\lambda=0.15$ along $[100]$ and $[111]$
directions: 
$U/W_B=-0.206$ [$E_B/W_b=0.0032$, $N_{eff}=44.3$] (circles);  
$U/W_B=-0.25$  [$E_B/W_b=0.0113$, $N_{eff}=11.2$] (rhombus);  
$U/W_B=-0.281$ [$E_B/W_b=0.0213$, $N_{eff}=6.05$] (squares);  
$U/W_B=-0.375$ [$E_B/W_b=0.0703$, $N_{eff}=2.38$] (triangles
pointing up);  
$U/W_B=-0.59$  [$E_B/W_b=0.2397$, $N_{eff}=1.35$] (triangles 
pointing down).  
}
\label{fig9}
\end{figure}
The dimensionless electron-phonon coupling 
constant $\lambda$ is defined to be 
$\lambda=\gamma^2/(4t\omega{\mbox{\scriptsize ph}})$.
It is clear that for the chosen model the parameter set 
to determine relation (\ref{nvse}) is  
${\cal P} = \left\{ U, \lambda \right\}$ including the 
potential of the attracting trap $U$ and the strength of the 
electron-phonon coupling $\lambda$.

To calculate the values  of $E_B=E_B(U,\lambda)$ and  
$N_{eff}=N_{eff}(U,\lambda)$ we used the direct space 
diagrammatic Monte Carlo (DSDMC) technique  \cite{DirDMC09}. 
Similar data can be can be obtained by the inhomogeneous
momentum average approximation method 
\cite{BercEPL10,BercPRL09} and the coherent 
basis states method \cite{CBS1,CBS2}. 
The data for $E_B=E_B(U,\lambda)$ and  
$N_{eff}=N_{eff}(U,\lambda)$ are presented in 
Fig.~\ref{fig8}a and \ref{fig8}b.  
The values of $N_{eff}$ were determined by relation
(\ref{Neff}) from the charge distribution in the trap which
was calculated by the DSDMC technique (see Fig.~\ref{fig9}).

To determine unambiguously the functional 
dependence $N_{eff}=N_{eff}(E_B)$
from the calculated relations $E_B=E_B(U,\lambda)$ and  
$N_{eff}=N_{eff}(U,\lambda)$ (Figs.~\ref{fig8}a and \ref{fig8}b) 
we must decide which of the two parameters, 
$\lambda$ and $U$, is fixed 
and which is responsible for the variation of 
the physical parameters of the traps: the binding energy
$E_B$ and the localization parameter $N_{eff}$.
A proper choice of the parameter responsible 
for the variation in the physical
properties of the traps determines relation (\ref{nvse})
fully and unambiguously. 
Since the thin film in pentacene TFTs consists solely of
pentacene molecules it is natural to assume that the value of
$\lambda$ is one and the same for the entire film and 
the spread of the physical parameters of the traps 
is caused by trapping potentials of different origins.
The trapping potentials of different origins, in turn, 
can be characterized by different strengths of 
the attractive potentials $U$.       

\subsection{Energy distribution of traps in pentacene TFTs}
\label{edis-tft}

To analyze the ESR spectrum of pentacene TFT 
we used the electron-phonon coupling constant $\lambda=0.15$,   
estimated from optical absorption experiments \cite{Bao07},
and the hopping amplitude $t=0.1$eV, obtained from 
band structure calculations \cite{Parisse07,OtherBand}. 
\begin{figure}[th]
        \includegraphics[scale=0.8]{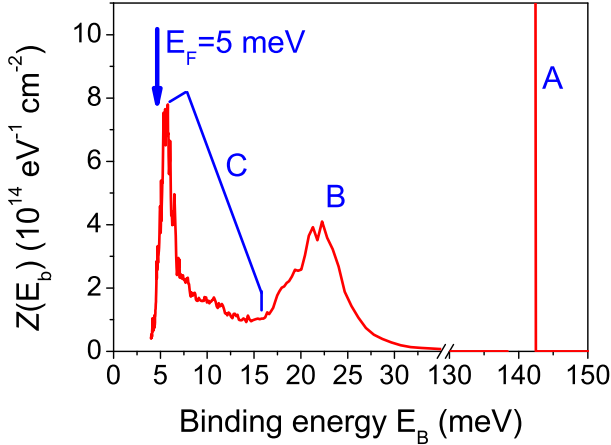}
\caption{(color online) 
Distribution of trap states in pentacene TFTs
as function of trap energy $E_B$ for gate voltage -200 V.}
\label{fig10}
\end{figure}
Figure \ref{fig10} shows the distribution ${\cal Z}(E_B)$ of the trapped carriers over the binding energies in TFT at the gate 
voltage -200 V.
We fixed $\lambda=0.15$, considered the dependence of $E_B$ 
(bold line in Fig.~\ref{fig8}a) and $N_{eff}$ 
(bold line in Fig.~\ref{fig8}b) on the attractive potential
$U$, and obtained $N_{eff}(E_B)$ 
(bold line in Fig.~\ref{fig8}c). 
Then, we used transformation (\ref{transf}). 
 
We found that the two discrete trap levels (A~and~B) 
peak at $140 \pm 50$ meV and $22 \pm 3$ meV, respectively.
The broad feature (C) at the gate voltage -200 V 
is distributed between 5 and 15 meV, as presented 
in Fig.~\ref{fig10}. 
The low-energy profile prompts the anticipation of 
tail states extending from just below the band edge, 
as has been discussed for amorphous semiconductors, 
while the states are partially occupied up to the 
Fermi level at around $E_B$ = 5~meV. 
These results are roughly consistent with the small activation 
energy of about 15 meV for the motional narrowing 
observed in [\onlinecite{MMH10}], 
and also with the potential fluctuations by 
atomic-force-microscope potentiometry \cite{Ohashi07}. 
We note that the distribution ${\cal Z}(E_B)$ gives relatively 
correct position of the trap levels, although the 
absolute value of the binding energy is rather 
model-dependent.
 
\section{Discussion}
\label{disc}

Weakly-localized in-gap states are expected to play crucial 
roles in the intrinsic charge transport along 
semiconductor/gate dielectric interfaces in organic 
transistors. In practice, temperature-independent mobility 
is often observed in devices with high mobility 
and highly-ordered molecular interfaces, which 
indicates that the Fermi energy is just below the 
band edge \cite{Hase09}. 

To date, various experimental
techniques have been used to investigate the 
interfacial trap density, such as deep-level transient 
spectroscopy (DLTS) \cite{Yang02}, photocurrent yield 
\cite{Lang04}, gate-bias stress \cite{Gol06}, 
and thermally-stimulated current \cite{Matsu07} experiments. 
However, the measurements are based on the charge 
transport, and it is strongly affected by the 
"extrinsic" potential barriers at grain boundaries 
and/or channel/electrode interfaces. In striking 
contrast, our method has a crucial advantage 
in its ability to disclose the spatial and energy 
distribution of shallow traps down to a few meV, 
based on a unique microscopic probe using electronic 
spins. In addition, the g tensor can be used to 
identify the molecular species around the trap sites. 
For the three types (A, B, and C) of trap 
states obtained, the g tensor should be common, considering 
the highly symmetric nature of the ESR spectra. 
This indicates that the trap states are extended 
over inherent pentacene molecules of regular 
orientation \cite{Matsu08,Maru06}. 
Of these, the deep discrete 
trap level (A) might be attributed to structural 
defects such as grain boundaries \cite{Verlaak07}. 
\begin{figure}
        \includegraphics[scale=1.1]{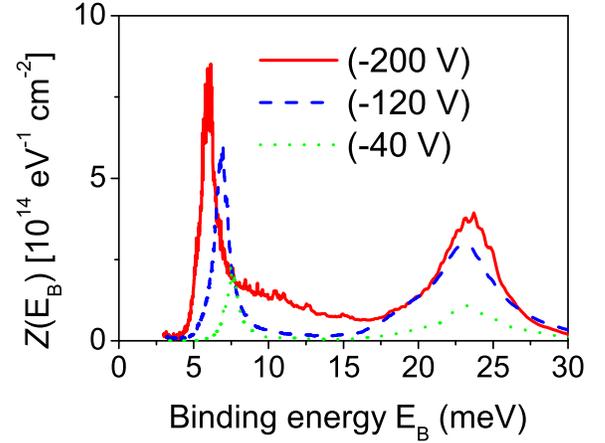}
\caption{(color online) 
Comparison of distributions 
of trap states in pentacene TFTs
as function of trap energy $E_B$ for gate 
voltage -200 V (solid line), 
-120 V (dashed line), and -40 V (dotted line).
}
\label{fig11}
\end{figure}

The shallow discrete 
level (B) and the broad feature (C) might be 
ascribed to small defects such as molecular 
sliding along the long axes of the molecules 
\cite{Kang05}, 
disorder induced by random dipoles in the 
amorphous gate dielectrics 
\cite{Veres03,Hou06}, thermal off-diagonal 
electronic disorder \cite{Troisi,Tro2010}, 
and the fluctuations of the band edge \cite{fluctu}. 
Notice that the regular orientations of the molecules 
are retained in the trap states as stated above. 
Although these assignments are rather speculative, 
we believe that further microscopic investigations 
based on this study will provide a 
comprehensive view of the weakly-localized 
in-gap states in organic transistors.
 
The lack of sensitivity of the strongly localized states with small
$N_{eff}$ to the gate voltage, which is obvious from
the physical point of view, is an indication of the stability
and reliability of the fine analysis of the 
ESR spectra.
For shallow states with large $N_{eff}$ and
small $E_B$, an increase in the gate voltage
adds low-energy states that participate in 
creation of ESR signal (Fig.~\ref{fig11}).
This shift of the border where states are visible 
to the ESR probe indicates that these states 
are filled as the gate voltage increases.
Hence, this behavior can be interpreted as a movement 
of the Fermi level which tends to zero energy as 
the bias voltage increases. 

It is important to mention that the sharp peak of $D(N_{eff})$
at $N_{eff}=1.54$ does not contradict the assumptions used
to derive the integral equations (\ref{conv1}) and (\ref{conv2}). 
The very essence of these equations implies that the contribution 
from each state with a given $N_{eff}$ is a  Gaussian ESR signal.
On the other hand, the signal at small $N_{eff}$ is a more 
complicated function with fine features (see e.g.
Fig.~\ref{fig1}a for $N_{eff}=1$).
The results for distribution $D(N_{eff})$ at small $N_{eff}$ 
can be unreliable. 
However, as can be seen in Fig.~\ref{fig1}c, the ESR signal for
reasonable parameters is close to 
Gaussian even at $N_{eff}=1.54$. 
Therefore, the results for the distribution of the traps $D(N_{eff})$
are valid even for small values of the localization 
parameter $N_{eff}$ .

\section{Conclusions}
\label{concl}

We have presented an unbiased analytical method for the
processing of high-precision electron spin resonance (ESR)
spectra, which allows us to obtain the distribution of trapped
carriers over the degree of localization and the binding energy.
The first step is a fine analysis of the shape of the
ESR spectra by the SOM, which
allows us to split the spectrum into multiple Gaussian
components each of which corresponds to a different spatial
extension of the trapped carriers. The second step is the
transformation of the distribution over the degree of
localization into a distribution over the binding energies via a
system-dependent relation between the binding energies and
the localization parameters of the trapped carriers. 
We have presented
and discussed the fundamental bases of the spectral analysis,
detailed algorithms for practical applications, and discussed the
robustness of the analysis to experimental noise. Although the
method can be applied to many systems,
we consider that it is most appropriate for 
ESR spectra of organic TFTs for the following reasons.
First, the channel materials are composed of regularly aligned
organic molecules that involve multiple degrees of freedom for
nuclear spin moments. This feature clearly justifies our basic
assumption that a single type of trap gives the
Gaussian lineshape of the ESR spectrum. Secondly, it is possible
to measure the high-precision ESR spectrum  because of the fairly
small spin-orbit interactions of organic materials. The
field-effect device structure also enables the control of the
carrier density without introducing any randomness in the
channel semiconductors.

Such a direct probe is quite unique in investigating the
microscopic carrier dynamics in the organic TFTs that have 
attracted considerable recent attention in the field of organic
electronics. We have shown that the trap
states in pentacene TFTs can be classified into three major
groups: deep trap states with a spatial extension of about 1.5
molecules (A), relatively shallow trap states that extend over
about 5 molecules (B), and shallower trap states that extend
over 6 to 20 molecules (C). These states respectively correspond
to deep and shallow trap states with binding energies of
140 meV (A) and 22 meV (B), with the broad feature
ranging from 5 to 15 meV (C). These shallow in-gap states are
crucial for understanding and improving the device performance
of organic TFTs.

\section{Acknowledgments}

A.S.M.\ was supported by RFBR 07-02-00067a. H.M. and T.H. are
supported by the Japan Science and Technology Agency (JST) through the Strategic Promotion of Innovative Research and Development Program (S-Innovation).
They are also partly supported by the Japan Society for the Promotion of Science (JSPS) through a Grant-in-Aid for Scientific Research from
the Funding Program for World-Leading Innovative R\&D on Science and Technology (FIRST program).

\end{document}